\documentclass[runningheads]{llncs}
\usepackage[T1]{fontenc}
\usepackage{graphicx}
\usepackage{xcolor}
\usepackage{siunitx}
\usepackage{textcomp} 
\usepackage{hyperref} 
\usepackage{color}

\definecolor{linkcolor}{RGB}{0,0,240}
\usepackage{booktabs}
\usepackage{tabularx}
\newcolumntype{Y}{>{\centering\arraybackslash}X}

\usepackage{amsmath}
\usepackage{overpic}
\usepackage{enumitem} 
\usepackage{overpic} 
\usepackage{color}
\usepackage[export]{adjustbox}
\usepackage{longtable}
\definecolor{turquoise}{cmyk}{0.65,0,0.1,0.3}
\definecolor{purple}{rgb}{0.65,0,0.65}
\definecolor{dark_green}{rgb}{0, 0.5, 0}
\definecolor{orange}{rgb}{0.8, 0.6, 0.2}
\definecolor{red}{rgb}{0.8, 0.2, 0.2}
\definecolor{darkred}{rgb}{0.6, 0.1, 0.05}
\definecolor{blueish}{rgb}{0.0, 0.3, .6}
\definecolor{light_gray}{rgb}{0.7, 0.7, .7}
\definecolor{pink}{rgb}{1, 0, 1}
\definecolor{greyblue}{rgb}{0.25, 0.25, 1}
\usepackage[normalem]{ulem}
\usepackage{tabularx}
\newcolumntype{Y}{>{\centering\arraybackslash}X}
\usepackage[T1]{fontenc}

\usepackage{blindtext}

\renewcommand{\paragraph}[1]{\vspace{1em}\noindent\textbf{#1}.}

\def\1{\mathbbm{1}}

\usepackage{amsfonts, cleveref}
\usepackage{bm}
\usepackage{bbm}

\usepackage{caption}
\usepackage{wrapfig}

\usepackage[normalem]{ulem}
\usepackage{booktabs}
\usepackage{multirow}

\begin{document}
\title{Self-Supervised Distillation of Legacy Rule-Based Methods for Enhanced EEG-Based Decision-Making}
\titlerunning{Enhancing Clinical Performance with Self-Supervised Distillation}
%
\author{%
Yipeng Zhang\inst{1}\orcidID{0000-0003-2869-4692}%
\index{Zhang, Yipeng} \and
Yuanyi Ding\inst{1}\orcidID{0009-0006-5566-675X}%
\index{Ding, Yuanyi} \and
Chenda Duan\inst{1}\orcidID{0009-0003-8652-3960}%
\index{Duan, Chenda} \and
Atsuro Daida\inst{2}\orcidID{0000-0003-1350-7700}%
\index{Daida, Atsuro} \and
Hiroki Nariai\inst{2}\orcidID{0000-0002-8318-2924}%
\index{Nariai, Hiroki} \and
Vwani Roychowdhury\inst{1}\orcidID{0000-0003-0832-6489}%
\index{Roychowdhury, Vwani}
}
\authorrunning{Y. Zhang et al.}
%
\institute{Samueli School of Engineering, UCLA, USA\\
\and
Mattel Children’s Hospital, David Geffen School of Medicine, UCLA, USA\\
\email{vwani@g.ucla.edu}
}
\maketitle              
\begin{abstract}
High-frequency oscillations (HFOs) in intracranial Electroencephalography (iEEG) are critical biomarkers for localizing the epileptogenic zone in epilepsy treatment. However, traditional rule-based detectors for HFOs suffer from unsatisfactory precision, producing false positives that require time-consuming manual review. Supervised machine learning approaches have been used to classify the detection results, yet they typically depend on labeled datasets, which are difficult to acquire due to the need for specialized expertise. Moreover, accurate labeling of HFOs is challenging due to low inter-rater reliability and inconsistent annotation practices across institutions. The lack of a clear consensus on what constitutes a pathological HFO further challenges supervised refinement approaches. To address this, we leverage the insight that legacy detectors reliably capture clinically relevant signals despite their relatively high false positive rates. We thus propose the Self-Supervised to Label Discovery (SS2LD) framework to refine the large set of candidate events generated by legacy detectors into a precise set of pathological HFOs. SS2LD employs a variational autoencoder (VAE) for morphological pre-training to learn meaningful latent representation of the detected events. These representations are clustered to derive weak supervision for pathological events. A classifier then uses this supervision to refine detection boundaries, trained on real and VAE-augmented data. Evaluated on large multi-institutional interictal iEEG datasets, SS2LD outperforms state-of-the-art methods. SS2LD offers a scalable, label-efficient, and clinically effective strategy to identify pathological HFOs using legacy detectors.
\end{abstract}
%

\section{Introduction}

Human-defined biomarkers encapsulate the expertise of medical professionals and play a pivotal role in medical diagnosis. In EEG-based clinical practice, neurophysiological biomarkers such as spikes, spindles, and high-frequency oscillations (HFOs) are commonly used. To detect these biomarkers computationally, numerous statistical methods have been developed, including the IDE spike detector \cite{uda2024normative}, the YASA spindle detector \cite{vallat2021open}, and HFO detection such as STE \cite{staba2004high} or MNI \cite{mni}. However, these rule-based approaches often suffer from low precision, generating false positives \cite{zweiphenning2022intraoperative}. Clinicians must manually review a large volume of detections, which is both time-consuming and prone to human error.

Machine learning approaches, especially supervised methods, have been employed to refine detection results \cite{jing2020development,ma2019automatic,zhang2024pyhfo} or to classify EEG segments produced by threshold-based “pseudo-detectors” \cite{xu2021bect}. While these methods can show good performance on their defined test set, they are typically derived from relatively small cohorts and heavily depend on accurately labeled training data, limiting their scalability and generalizability. Annotating such events requires specialized medical knowledge, and clinical studies have shown low inter-rater reliability for these biomarkers \cite{nariai2018interrater,spring2017interrater,scheuer2017spike}; standardization of annotation practices across institutions is also lacking \cite{johnson2023interictal,kini2019virtual}. Weakly supervised methods~\cite{BrainComm,MONSOOR2023} reduce annotation demands, but require complex loss tuning, preventing clinical use.

Nonetheless, earlier studies have predominantly overlooked a critical insight: rule-based biomarker detectors, developed over many years of clinical research, have already been successfully deployed in clinical practice—even after manual inspection and refinement. These detectors typically maintain a very high recall rate \cite{zelmann2012comparison,warby2014sleep,von2012automatic}. In other words, although the detected events may include many false positives, these legacy detectors reliably capture a large share of signals of clinical interest. Furthermore, since the verification of these detected events is often performed through visual inspection of EEG tracing, the definition of clinically meaningful characterization is reflected in the EEG morphology. This insight suggests that clinically relevant biomarkers can be distilled from the output of the rule-based detectors in a self-supervised manner—particularly in scenarios where expert annotations are limited in large datasets or the precise definition of clinically meaningful events remains elusive. 

Motivated by this insight, we leverage legacy detectors to identify clinically meaningful events (pathological) without requiring human labeling. Specifically, we focus on high-frequency oscillations (HFOs), a crucial intracranial EEG (iEEG) biomarker used to localize the epileptogenic zone from interictal (in-between seizures) recordings in epilepsy treatment \cite{jacobs2010high}. Importantly, a clear consensus on the definition of pathological HFO is not yet established \cite{zijlmans2012high}.

To address this challenge, we propose a Self-Supervised to Label Discovery (SS2LD) framework that automatically distills pathological HFOs from the large volume of events detected by legacy HFO detectors. Specifically, we first perform morphological pre-training using a variational autoencoder (VAE) to learn morphological representations of detector-identified events. Leveraging the model’s ability to encode distinct morphological features, we then use the VAE’s latent representations to cluster these events, extracting weak supervision corresponding to pathological events. Finally, we train a classifier on the pre-trained VAE for both feature extraction and data augmentation. This classifier takes latent codes from the VAE encoder and is trained using both real and VAE-augmented data derived from the discovered weak supervision, thus refining classification boundaries. SS2LD distills the events generated by legacy detectors and improves clinical decision-making. Our contributions are summarized as follows:
\begin{itemize}[leftmargin=*]
    \item We propose the SS2LD framework that uses VAE-based morphological pre-training to automatically discover pathological HFO events without human labels and exploits these discoveries as weak supervision for classification.
    \item We employ the VAE’s generative capability to augment training data, enhancing the classifier’s robustness in distilling pathological HFOs.
    \item SS2LD is evaluated on large multi-institutional iEEG datasets, demonstrating superior clinical performance compared to existing state-of-the-art methods and underscoring its potential on improving clinical decision-making.
\end{itemize}

\section{Methods}
\subsection{High-Frequency Oscillations in Intracranial EEG Recordings}
For each subject \( k \) in an iEEG cohort, the recording \( \mathcal{X}^k \) comprises multiple channels, indexed by \( c \), selected clinically to target probable epileptogenic regions. An HFO event \( e_i \) in channel \( c \) of \( \mathcal{X}^k \) is defined by its start time \( s_i \) and end time \( e_i \), such that \( e_i = \mathcal{X}^k_c[s_i : e_i] \). These segments may reflect pathological (epileptogenic) or non-pathological (normal) activity. Since HFO durations \( e_i - s_i \) vary, we construct a fixed-length window \( w_i = \mathcal{X}^k_c\bigl[\frac{s_i + e_i}{2} - L : \frac{s_i + e_i}{2} + L\bigr] \), centered at the event’s midpoint, to ensure consistent input sizes for neural networks.
\begin{figure}[!h]
    \centering
    \includegraphics[width=0.85\textwidth]{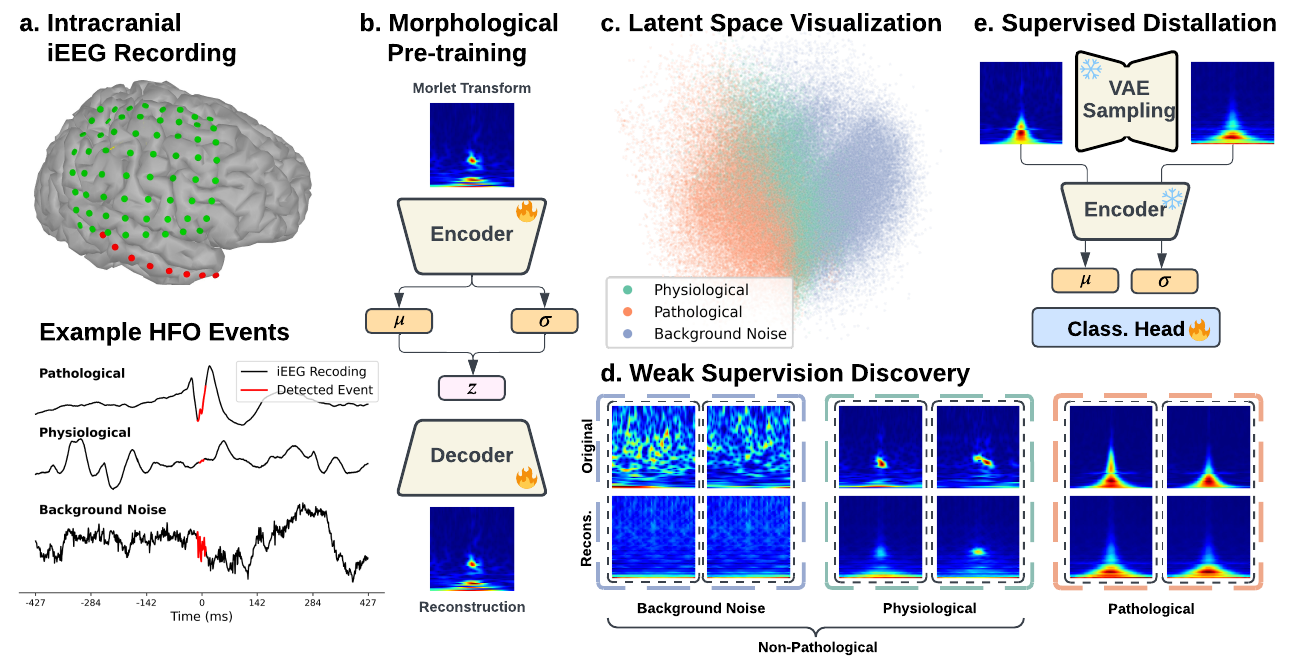}
    \caption{\textbf{Overview of the Proposed SS2LD Framework.} 
    (a) HFO events are detected in  interictal (in-between seizures) intracranial EEG recordings using rule-based detectors, including pathological, physiological, and background noise events. 
    (b) A VAE is pre-trained to reconstruct HFO morphological features, derived via Morlet Transform, with the encoder mapping inputs to latent distributions and the decoder reconstructing sampled latent codes. 
    (c) Latent representations are projected by PCA, with a color-coded 2D plot representing clustered event types. 
    (d) Hierarchical clustering identifies distinct morphological groups, enabling weak supervision for classifying pathological events. 
    (e) A classifier is trained on both original latent codes (\(\mu\)) and VAE-generated samples (\(\hat{\mu}\)) to further refine the decision boundary.}
    \label{fig:fig1}
\end{figure}

\subsection{Self-supvervised Pre-Training}
\subsubsection{Variational Autoencoder:} We train a VAE to capture HFOs' morphological features. The VAE includes a \emph{Time-Frequency Morphological Encoder} \( \mathcal{E}_{\theta} \) and a \emph{Morphological Decoder} \( \mathcal{D}_{\gamma} \). The encoder \( \mathcal{E}_{\theta} \) transforms an HFO event \( e_i \), represented as a fixed-length signal \( w_i \), into a time-frequency representation \( \mathcal{W}_i = \text{MorletTransform}(w_i) \), then projects it into a lower-dimensional latent space, yielding a distribution parameterized by \( \mu_{\theta}(w_i) \) and \( \Sigma_{\theta}(w_i) \). A latent code \( \mathbf{z}_i \sim \mathcal{N}\bigl(\mu_{\theta}(w_i), \Sigma_{\theta}(w_i)\bigr) \) is sampled, and the decoder \( \mathcal{D}_{\gamma} \) reconstructs \( \hat{\mathcal{W}}_i = \mathcal{D}_{\gamma}(\mathbf{z}_i) \), reconstructing the HFO’s morphological feature.

\subsubsection{Encouraging Morphological Understanding:}
To measure discrepancies between the real and reconstructed representations, we employ the perceptual loss \cite{johnson2016perceptual} \(\mathcal{L}_{\text{perceptual}}\), which uses the L\(_2\) distance between feature maps extracted from different layers of a pre-trained VGG16 network \(\phi(\cdot)\). Specifically, \(\mathcal{L}_{\text{perceptual}} = \sum_{l}\|\phi_l(\mathcal{W}_i) - \phi_l(\hat{\mathcal{W}}_i)\|_2^2\). To further disentangle the latent space and enhance the morphological expressiveness, we adopt a learnable \(\beta\)-VAE strategy. We introduce learnable  \(\beta\) on the original VAE loss function and we dynamically adjust \(\beta\) to balance between reconstruction and disentanglement of latent representations throughout training. Specifically, in each minibatch, the \(\beta\) is updated by: \(\beta_{\text{new}} = \beta + \beta_{\text{lr}} \cdot (\mathbb{E}[\mathcal{L}_{\text{KL}}] - \mathbb{E}[\mathcal{L}_{\text{perceptual}}])\), where \(\beta_{\text{lr}}\) is a learning rate for \(\beta\). Finally, the pre-train objective is: \(\mathcal{L_\text{pretrain}} = (1- \beta) \mathcal{L}_{\text{perceptual}} + \beta\,\mathcal{L}_{\text{KL}}\) (\(\beta\) $\in [0,1]$). Minimizing this loss encourages precise reconstructions and a disentangled, regularized latent representation.


\subsection{Weak Label Discovery from Structured Latent Space} After pre-training, we aim to gather meaningful insight into the latent representation by visualizing the input signals, their corresponding reconstructed signals, and the associated latent codes within the latent space. In Figure~\ref{fig:fig1}, we observe that the entire latent space could be broadly clustered into three morphologies, each with its own clinical significance: (1) pathological, (2) physiological, and (3) background noise. Furthermore, the model's reconstruction ability for background noise is limited (Figure~\ref{fig:fig1}d), likely due to the diverse morphologies of such events. To provide weak supervision, we conduct hierarchical clustering in the latent space. We first fit a k-means cluster with \(k=2\) across the entire latent space to separate background noise from HFO events, selecting the cluster with the higher reconstruction loss as the background class. Next, we cluster the remaining HFO events (again with \(k=2\)) to distinguish pathological from physiological events. We label the cluster predominantly located within the resected region as pathological. Finally, we assign a weak supervision label \(l = 1\) to pathological events and \(l = 0\) otherwise (physiological and background noise).

\subsection{VAE-Based Augmentation for Biomarker Distillation}  The latent-space visualization suggests that distinct morphological characteristics of the biomarkers are separated within this space. However, since k-means does not account for complex latent geometries, the class boundaries may not conform well to the actual distribution of biomarkers. We, thus, freeze the VAE parameters and train a classification head \(\mathcal{F}_{\psi}\) on top of the encoder to predict pathological events. By leveraging and distilling the weak supervision introduced by k-means, this training step refines the decision boundary. Furthermore, to improve generalization, we augment the training with VAE-generated surrogates: for each \( \mathcal{W}_i \), we sample \( \hat{\mathcal{W}}_i \). We then encode both \( \mathcal{W}_i \) and \( \hat{\mathcal{W}}_i \) to latent vectors \( \mu_i \) and \( \hat{\mu}_i \). Assuming similar morphology implies the same category, the classifier loss is \( \mathcal{L}_{\text{classifier}} = \text{BCE}\bigl(\mathcal{F}_{\psi}(\mu_i), l_i\bigr) + \text{BCE}\bigl(\mathcal{F}_{\psi}(\hat{\mu}_i), l_i\bigr) \), ensuring consistent predictions across original and surrogate samples.
\section{Results}

\subsection{Dataset}
We conduct experiments using the Open iEEG dataset~\cite{ds005398:1.0.1}, which comprises recordings from 185 epilepsy patients from two institutions (UCLA \& Detroit). Surgical outcomes (post-resection seizure freedom) are available for 162 patients (113 seizure-free). Each electrode (channel) is clinically annotated as resected or non-resected, with a total of 686,410 detected HFOs (from two kinds of HFO detector, STE and MNI). We randomly shuffle the subjects and conduct subject-wise fivefold cross-validation. We also extend our experiment using Zurich iEEG dataset~\cite{ds003498:1.1.1}, which includes 20 epilepsy patients. Following \cite{zurich}, we apply a bipolar montage to the EEG recordings, yielding 15 patients (10 seizure-free) with valid resection annotations (excluding five due to annotation errors of resection margin). We detect 97,511 HFOs (STE \& MNI) using the HFO detection parameters in \cite{zhang2024pyhfo}. Since each subject in the Zurich iEEG dataset has multiple runs, we treat all detected events across runs as a single set.

\subsection{Implementation Details}
We implement the VAE using a convolutional neural network with residual layers as the backbone, setting the latent dimension to 16. Each HFO event is represented in a fixed window, producing a time-frequency plot spanning 10–290 Hz over 570 ms, then resized to \(64 \times 64\). For the Open iEEG dataset, we divide each fold into training (n = 119), validation (n = 30), and test sets (n = 36), ensuring balanced institutional representation (UCLA/Detroit). The VAE is trained with stratified sampling capping samples per subject at 2,500 per epoch. Pre-training occurs from scratch using the Adam optimizer (\(lr = 10^{-3}\), weight decay \(\lambda = 10^{-5}\)) for 100 epochs with a batch size of 512, initializing \(\beta_{\text{init}} = 1\) and \(lr_{\beta} = 10^{-4}\). For classifier training, we use all training set events per epoch, training with Adam (\(lr = 3 \times 10^{-4}\), \(\lambda = 10^{-5}\)) and batch size of 4096 for nine epochs. All experiments are performed on NVIDIA A6000 GPUs.

\subsection{Evaluation Metrics}
Since there are no explicit ground-truth label for individual HFO, we evaluated our biomarker classifier using clinical evidence. We introduce two metrics to evaluate the effectiveness of the predicted pathological HFOs: ability to predict post-operative seizure outcome and classification specificity.

\subsubsection{Surgical Outcome Prediction:}  
Predicting surgical outcomes is one of the most promising ways to validate the clinical effectiveness of a biomarker, as it provides a real-world assessment of its prognostic value in guiding epilepsy treatment. A key feature used in this validation is the Resection Ratio. The Resection Ratio (\( RR^k \)) measures the proportion of pathological HFOs located in surgically resected channels for patient \( k \), which can be  defined as \( RR^k = \frac{\text{Number of pathological HFOs in resected channels}}{\text{Total number of pathological HFOs}} \), a value of \( RR^k = 1 \) indicates complete resection of pathological HFOs, while \( RR^k = 0 \) implies no overlap. To predict surgical outcomes (success or failure), we train a logistic regression model with balanced label weights using \( RR^k \) on the training and validation sets, then evaluate it on the test set and Zurich iEEG dataset. Given the class imbalance (\(\sim\)2/3 success), we report both accuracy (ACC) and F1 scores across five folds.

\subsubsection{Classification Specificity:}  
We leverage another evidence to validate the effectiveness of the predicted pathological HFO. If a patient remains seizure-free after resection, the preserved brain region should contain few or no pathological HFO. We compute the specificity for each surgical success patient \( k \) as \( \text{Specificity}^k = \frac{\text{Number of non-pathological HFO in the preserved region}}{\text{Total number of HFO detected in the preserved region}} \), where a higher specificity indicates better classification performance. The overall specificity (SPEC) is then defined by averaging across all surgical success patients (across five folds), given by \( \text{SPEC} = \frac{1}{|S|} \sum_{k \in S} \text{Specificity}^k \), where \( S \) is the set of successful cases.

\begin{table}[ht]
\centering
\caption{Performance comparison in Open iEEG and Zurich iEEG datasets.}
\label{tab:tab1}
\begin{tabularx}{\textwidth}{@{}l|YYc|YYc@{}} 
\toprule
\multirow{2}{*}{} & \multicolumn{3}{c|}{\textbf{Open iEEG Dataset}} & \multicolumn{3}{c}{\textbf{Zurich iEEG Dataset}} \\ 
\cmidrule(lr){2-7} 
      & \textbf{ACC}  & \textbf{F1} & \textbf{SPEC} & \textbf{ACC} & \textbf{F1} & \textbf{SPEC} \\ 
\midrule
eHFO \cite{MONSOOR2023} & $0.538 \pm 0.119$ & $0.397 \pm 0.069$ & 0.543 & $0.533\pm 0.000$ & $0.364 \pm 0.000$ & 0.745 \\ 
spkHFO \cite{zhang2024pyhfo} & $0.594 \pm 0.129$ & $0.434 \pm 0.070$ & 0.703 & $0.600 \pm 0.000$ & $0.500 \pm 0.000$ & 0.819 \\ 
\midrule
SS2LD & \textbf{0.612} $\pm 0.131$ & \textbf{0.464} $\pm 0.069$ & \textbf{0.749} & $\textbf{0.640} \pm 0.037$ & $\textbf{0.583} \pm 0.050$ & \textbf{0.909} \\ 
\bottomrule
\end{tabularx}
\end{table}

\subsection{Comparison with State-of-the-Art Methods}
We compare SS2LD against two state-of-the-art pipelines for classifying pathological HFOs, leveraging their publicly available models. These pathological HFOs are defined as: \textbf{(1) Epileptogenic HFO (eHFO)}~\cite{MONSOOR2023}, identified via weakly supervised learning rooted in clinical evidence; and \textbf{(2) HFO with Spike Discharge (spkHFO)}~\cite{zhang2024pyhfo}, trained on human-annotated events informed by clinical expertise. Table~\ref{tab:tab1} presents the performance of SS2LD alongside other pipelines, reporting the averaged ACC and F1 scores across five folds, along with their standard deviations (STD). Specificity is averaged across all patients for the Open iEEG dataset. For the Zurich iEEG dataset, since SS2LD is trained using five-fold cross-validation, specificity is first averaged across patients and then across the five folds. For~\cite{MONSOOR2023,zhang2024pyhfo}, the classifier is trained on distinct patient subsets from the Open iEEG dataset over five folds; however, decision boundary variations do not impact performance on the Zurich iEEG dataset, yielding an STD of zero for ACC and F1 scores. Nevertheless, SS2LD outperforms both in seizure outcome prediction and specificity across all datasets.
\subsubsection{Ablation Studies:}

We present ablation studies on key components of the proposed SS2LD:
(1) \textit{Weak Supervision}, using weak supervision provided from k-means clustering;  
(2) \textit{Supervised Distillation}, training the classifier with supervised distillation (SD) but without VAE augmentation (AUG).  Additionally, we investigate the impact of latent space dimensions (DIM = 8, 32, and 64). A latent dimension of 8 appears overly restrictive, limiting the model’s ability to capture critical details of pathological events. While SS2LD achieves higher specificity with latent dimensions above 16, the denser latent details can lead to overfitting and overly conservative predictions (favoring non-pathological labels), potentially reducing its clinical utility. Table~\ref{tab:tab2} highlights the effectiveness of our proposed modules and the selection of latent dimensions.
\begin{table}[ht]
\centering
\caption{Ablation study results for SS2LD on the Open iEEG dataset.}
\label{tab:tab2}

\begin{tabularx}{\textwidth}{@{}c|cc|c|YYc@{}}
\toprule
 & \textbf{SD} & \textbf{AUG} & \textbf{DIM} & \textbf{ACC}  & \textbf{F1} & \textbf{SPEC}\\ 
\midrule
Weak Supervision         &             &             & 16    & $0.557 \pm 0.101$ & $0.423 \pm 0.066$ & 0.739 \\ 
Supervised Distillation     & \checkmark  &             & 16    & $0.568 \pm 0.116$ & $0.433 \pm 0.058$ & 0.715\\
\midrule
\multirow{4}{*}{SS2LD}
   & \checkmark & \checkmark & 8   & 0.562 $\pm 0.134$ & 0.426 $\pm 0.065$ & 0.736\\ 
   & \checkmark & \checkmark & 16  & \textbf{0.612} $\pm 0.108$ & \textbf{0.464} $\pm 0.063$ & 0.749\\
   & \checkmark & \checkmark & 32  & 0.593 $\pm 0.128$ & 0.442 $\pm 0.065$ & 0.761\\ 
   & \checkmark & \checkmark & 64  & 0.574 $\pm 0.126$ & 0.431 $\pm 0.061$ & 0.752\\ 
\bottomrule
\end{tabularx}


\end{table}
\begin{figure}[!h]
    \centering
    \includegraphics[width=0.85\textwidth]{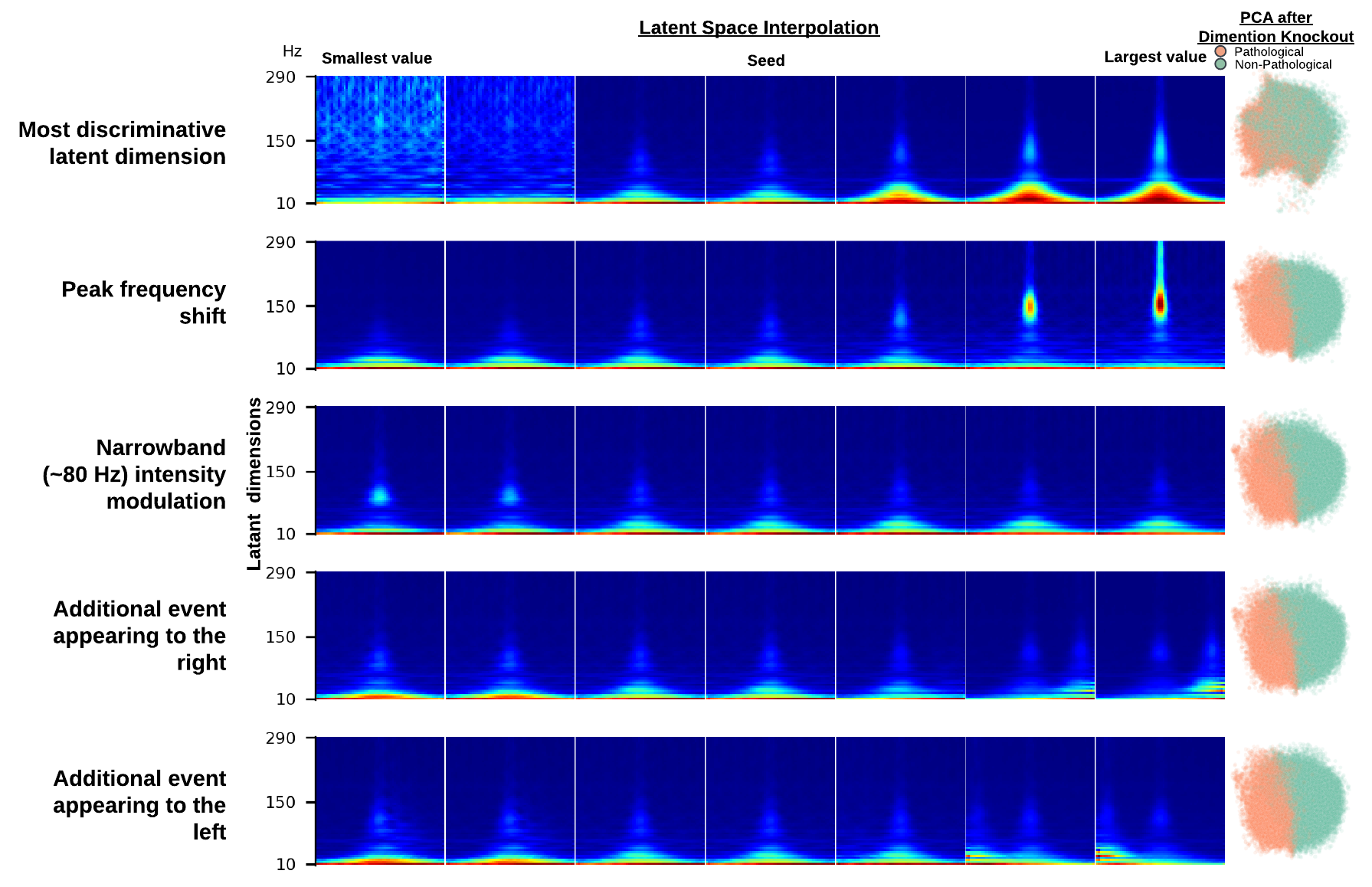}
    \caption{\textbf{Latent Dimension Interpolation and Dimension Knockout:} We visualize the interpolation results focusing on the top 5 most expressive latent dimensions in one example fold.
    \textbf{(Left)} Each row corresponds to a specific latent dimension, annotated with its inferred role in encoding HFO morphology, including variations in spectral intensity across frequency bands (e.g.,~80 Hz), peak frequency shifts, and the presence of additional events. The central spectrogram represents the original unperturbed "seed", derived from the mean latent code of predicted pathological events. Notably, the first dimension corresponds to background noise to pathological events. 
    \textbf{(Right)} PCA is performed after knocking out each latent dimension to assess its influence on the classifier’s decision boundary. Each scatter plot visualizes the distribution of latent embeddings post-knockout, with colors indicating classifier-predicted pathological (orange) or non-pathological (green) events. Overall, the first dimension contributes the most to the final decision boundary.
    }
    \label{fig:fig2}
\end{figure}

\subsection{Neurophysiological Meaningful Latent Space}
To investigate the neurophysiological role of each latent dimension, we interpolate the latent space using the mean latent code of predicted pathological events as a seed. Each dimension is adjusted from its smallest (0.001 percentile) to largest (0.999 percentile) value across all latent codes. The VAE decoder then transforms these interpolated latent codes into their corresponding time-frequency representations, enabling us to visualize how variations in each dimension affect the encoded neurophysiological features. Furthermore, to evaluate each dimension's influence on the decision boundary, we conduct latent knockout by setting a dimension to zero, recalculating the PCA, and color-coding results by classifier predictions. Increased mixing of pathological and non-pathological predictions suggests a stronger contribution to the decision boundary. Figure~\ref{fig:fig2} illustrates how distinct latent dimensions encode unique HFO morphological features and affect the decision boundary, demonstrating the meaningful latent space that VAE learned.

\newpage

%
%
\bibliographystyle{splncs04}
\bibliography{main}

\begin{thebibliography}{10}
\providecommand{\url}[1]{\texttt{#1}}
\providecommand{\urlprefix}{URL }
\providecommand{\doi}[1]{https://doi.org/#1}

\bibitem{von2012automatic}
von Ellenrieder, N., Andrade-Valen{\c{c}}a, L.P., Dubeau, F., Gotman, J.: Automatic detection of fast oscillations (40--200 hz) in scalp eeg recordings. Clinical Neurophysiology  \textbf{123}(4),  670--680 (2012)

\bibitem{zurich}
Fedele, T., Burnos, S., Boran, E., Krayenb{\"u}hl, N., Hilfiker, P., Grunwald, T., Sarnthein, J.: Resection of high frequency oscillations predicts seizure outcome in the individual patient. Scientific Reports  \textbf{7}(1),  13836 (2017)

\bibitem{jacobs2010high}
Jacobs, J., Zijlmans, M., Zelmann, R., Chatillon, C.{\'E}., Hall, J., Olivier, A., Dubeau, F., Gotman, J.: High-frequency electroencephalographic oscillations correlate with outcome of epilepsy surgery. Annals of Neurology: Official Journal of the American Neurological Association and the Child Neurology Society  \textbf{67}(2),  209--220 (2010)

\bibitem{jing2020development}
Jing, J., Sun, H., Kim, J.A., Herlopian, A., Karakis, I., Ng, M., Halford, J.J., Maus, D., Chan, F., Dolatshahi, M., et~al.: Development of expert-level automated detection of epileptiform discharges during electroencephalogram interpretation. JAMA neurology  \textbf{77}(1),  103--108 (2020)

\bibitem{johnson2023interictal}
Johnson, G.W., Doss, D.J., Morgan, V.L., Paulo, D.L., Cai, L.Y., Shless, J.S., Negi, A.S., Gummadavelli, A., Kang, H., Reddy, S.B., et~al.: The interictal suppression hypothesis in focal epilepsy: network-level supporting evidence. Brain  \textbf{146}(7),  2828--2845 (2023)

\bibitem{johnson2016perceptual}
Johnson, J., Alahi, A., Fei-Fei, L.: Perceptual losses for real-time style transfer and super-resolution. In: Computer Vision--ECCV 2016: 14th European Conference, Amsterdam, The Netherlands, October 11-14, 2016, Proceedings, Part II 14. pp. 694--711. Springer (2016)

\bibitem{kini2019virtual}
Kini, L.G., Bernabei, J.M., Mikhail, F., Hadar, P., Shah, P., Khambhati, A.N., Oechsel, K., Archer, R., Boccanfuso, J., Conrad, E., et~al.: Virtual resection predicts surgical outcome for drug-resistant epilepsy. Brain  \textbf{142}(12),  3892--3905 (2019)

\bibitem{ma2019automatic}
Ma, K., Lai, D., Chen, Z., Zeng, Z., Zhang, X., Chen, W., Zhang, H.: Automatic detection of high frequency oscillations (80-500hz) based on convolutional neural network in human intracerebral electroencephalogram. In: 2019 41st Annual International Conference of the IEEE Engineering in Medicine and Biology Society (EMBC). pp. 5133--5136. IEEE (2019)

\bibitem{MONSOOR2023}
Monsoor, T., Zhang, Y., Daida, A., Oana, S., Lu, Q., Hussain, S.A., Fallah, A., Sankar, R., Staba, R.J., Speier, W., Roychowdhury, V., Nariai, H.: Optimizing detection and deep learning-based classification of pathological high-frequency oscillations in epilepsy. Clinical Neurophysiology  (2023). \doi{https://doi.org/10.1016/j.clinph.2023.07.012}, \url{https://www.sciencedirect.com/science/article/pii/S1388245723006971}

\bibitem{nariai2018interrater}
Nariai, H., Wu, J.Y., Bernardo, D., Fallah, A., Sankar, R., Hussain, S.A.: Interrater reliability in visual identification of interictal high-frequency oscillations on electrocorticography and scalp eeg. Epilepsia open  \textbf{3},  127--132 (2018)

\bibitem{scheuer2017spike}
Scheuer, M.L., Bagic, A., Wilson, S.B.: Spike detection: Inter-reader agreement and a statistical turing test on a large data set. Clinical Neurophysiology  \textbf{128}(1),  243--250 (2017)

\bibitem{spring2017interrater}
Spring, A.M., Pittman, D.J., Aghakhani, Y., Jirsch, J., Pillay, N., Bello-Espinosa, L.E., Josephson, C., Federico, P.: Interrater reliability of visually evaluated high frequency oscillations. Clinical Neurophysiology  \textbf{128}(3),  433--441 (2017)

\bibitem{staba2004high}
Staba, R.J., Wilson, C.L., Bragin, A., Jhung, D., Fried, I., Engel~Jr, J.: High-frequency oscillations recorded in human medial temporal lobe during sleep. Annals of neurology  \textbf{56}(1),  108--115 (2004)

\bibitem{ds003498:1.1.1}
T, F., N, K., P, H., Li, A., J., S.: "interictal ieeg during slow-wave sleep with hfo markings" (2023). \doi{doi:10.18112/openneuro.ds003498.v1.1.1}

\bibitem{uda2024normative}
Uda, H., Kuroda, N., Firestone, E., Ueda, R., Sakakura, K., Kitazawa, Y., Choromanski, D., Cools, M., Luat, A.F., Asano, E.: Normative atlases of high-frequency oscillation and spike rates under sevoflurane anesthesia. Clinical Neurophysiology  \textbf{167},  117--130 (2024)

\bibitem{vallat2021open}
Vallat, R., Walker, M.P.: An open-source, high-performance tool for automated sleep staging. Elife  \textbf{10},  e70092 (2021)

\bibitem{warby2014sleep}
Warby, S.C., Wendt, S.L., Welinder, P., Munk, E.G., Carrillo, O., Sorensen, H.B., Jennum, P., Peppard, P.E., Perona, P., Mignot, E.: Sleep-spindle detection: crowdsourcing and evaluating performance of experts, non-experts and automated methods. Nature methods  \textbf{11}(4),  385--392 (2014)

\bibitem{xu2021bect}
Xu, Z., Wang, T., Cao, J., Bao, Z., Jiang, T., Gao, F.: Bect spike detection based on novel eeg sequence features and lstm algorithms. IEEE Transactions on Neural Systems and Rehabilitation Engineering  \textbf{29},  1734--1743 (2021)

\bibitem{zelmann2012comparison}
Zelmann, R., Mari, F., Jacobs, J., Zijlmans, M., Dubeau, F., Gotman, J.: A comparison between detectors of high frequency oscillations. Clinical Neurophysiology  \textbf{123}(1),  106--116 (2012)

\bibitem{mni}
Zelmann, R., Mari, F., Jacobs, J., Zijlmans, M., Chander, R., Gotman, J.: Automatic detector of high frequency oscillations for human recordings with macroelectrodes. In: 2010 Annual International Conference of the IEEE Engineering in Medicine and Biology. pp. 2329--2333. IEEE (2010)

\bibitem{ds005398:1.0.1}
Zhang, Y., Daida, A., Liu, L., Kuroda, N., Ding, Y., Oana, S., Monsoor, T., Duan, C., Hussain, S.A., Qiao, J.X., Salamon, N., Fallah, A., Sim, M.S., Sankar, R., Staba, R.J., Jr., J.E., Asano, E., Roychowdhury, V., Nariai, H.: "open ieeg dataset" (2024). \doi{doi:10.18112/openneuro.ds005398.v1.0.1}

\bibitem{zhang2024pyhfo}
Zhang, Y., Liu, L., Ding, Y., Chen, X., Monsoor, T., Daida, A., Oana, S., Hussain, S., Sankar, R., Fallah, A., et~al.: Pyhfo: lightweight deep learning-powered end-to-end high-frequency oscillations analysis application. Journal of Neural Engineering  \textbf{21}(3),  036023 (2024)

\bibitem{BrainComm}
Zhang, Y., Lu, Q., Monsoor, T., Hussain, S.A., Qiao, J.X., Salamon, N., Fallah, A., Sim, M.S., Asano, E., Sankar, R., et~al.: Refining epileptogenic high-frequency oscillations using deep learning: a reverse engineering approach. Brain communications  \textbf{4}(1),  fcab267 (2022)

\bibitem{zijlmans2012high}
Zijlmans, M., Jiruska, P., Zelmann, R., Leijten, F.S., Jefferys, J.G., Gotman, J.: High-frequency oscillations as a new biomarker in epilepsy. Annals of neurology  \textbf{71}(2),  169--178 (2012)

\bibitem{zweiphenning2022intraoperative}
Zweiphenning, W., van't Klooster, M.A., van Klink, N.E., Leijten, F.S., Ferrier, C.H., Gebbink, T., Huiskamp, G., van Zandvoort, M.J., van Schooneveld, M.M., Bourez, M., et~al.: Intraoperative electrocorticography using high-frequency oscillations or spikes to tailor epilepsy surgery in the netherlands (the hfo trial): a randomised, single-blind, adaptive non-inferiority trial. The Lancet Neurology  \textbf{21}(11),  982--993 (2022)

\end{thebibliography}
\end{document}